\documentclass[twocolumn,prl,amssymb,epsfig,aps]{revtex4}
\usepackage{epsfig}
\usepackage{ifsym}
\usepackage{wasysym}

\begin{document}
\textheight 226.5mm
\title{\Large
{\sf Structures and topological transitions of hydrocarbon films on quasicrystalline surfaces}}
\author{
Wahyu Setyawan$^{1}$, Renee D. Diehl$^{2}$, and Stefano Curtarolo$^{1,\star}$}
\affiliation{
$^1$Department of Mechanical Engineering and Materials Science Duke University Durham, NC 27708 \\
$^2$Department of Physics and Materials Research Institute Penn State University University Park, PA 16801\\
$^\star${corresponding author, e-mail: stefano@duke.edu}
}
\date{\today}
\begin{abstract}
  Lubricants can affect quasicrystalline coatings surfaces by
  modifying commensurability of the interfaces.  We report results of
  the first computer simulation studies of physically adsorbed
  hydrocarbons on a quasicrystalline surface: methane, propane, and
  benzene on decagonal Al-Ni-Co.  The grand canonical Monte Carlo
  method is employed, using novel Embedded Atom Method potentials
  generated from {\it ab initio} calculations, and standard
  hydrocarbon interactions.  The resulting adsorption isotherms and
  calculated structures show the films' evolution from submonolayer to
  condensation.  We discover the presence and absence of the 5- to
  6-fold topological transition, for benzene and methane,
  respectively, in agreement with a previsouly formulated
  phenomenological rule based on adsorbate-substrate size mismatch.
\end{abstract}
\maketitle
% PACS 61.44.Br  68.43.Hn   68.55.Ac

Friction can become vanishingly small between incommensurate
interfaces (superlubricity) \cite{Hirano90,Dienwiebel04}.  The
aperiodic structure of quasicrystals surfaces makes them the ideal
candidates for this phenomenon.  Indeed, the special frictional
properties of quasicrystals have been confirmed in a series of
experiments using atomic force microscopy on single grain
quasicrystals \cite{Park1,Park2,Park3,Park4,Park5,Park6}.  Although
its origin is not yet completely understood, the evidence suggests
that poor coupling of phonons at the interfaces may play a major role
\cite{Park4,Park5,Park6}.  In addition, there is also evidence that
oxidation decreases the friction even further \cite{Park2,Dubois06}.
In fact, even before these single crystal experiments were performed,
experiments on quasicrystal coatings in air did show low-friction
behavior \cite{Dubois1}, leading to applications involving moving
machine parts and non-stick cookware \cite{Dubois_book}.

Since the low friction of quasicrystals is clearly related to their
structure, the interaction of the lubricant with the quasicrystal and
the structure of the lubricating film is particularly important. In
most applications involving machine parts, additional lubricant would
be needed to address the frictional contributions of grain boundaries
and asperities.  Studies of metal and rare gas adsorption on
quasicrystals indicate that both periodic and aperiodic structures can
occur in thin films
\cite{Fournee_2005,Sharma_2007,Ferralis_2004,XeQCPRL,XeQCPRB,XeQCJPCM}.
However, little is known about the interaction of hydrocarbons with
quasicrystal surfaces \cite{Park3,Hoeft_2006,McGrath_2002}.

In the present manuscript, we explore the effect of structural and
symmetry mismatches on the ordering of hydrocarbons by evaluating the
nature of hydrocarbon adsorption (methane, propane, and benzene) on a
quasicrystalline decagonal surface, namely, the tenfold surface of
Al$_{73}$Ni$_{10}$Co$_{17}$.  The simulations are performed using
grand canonical Monte Carlo method (GCMC), with which we have
extensive experience on smooth \cite{ref14,ref15,refSMOOTH} and
corrugated substrates
\cite{refHETERO,XeQCPRL,XeQCPRB,XeQCJPCM,XeQCPHILMAG,Renee08}.  Using
the GCMC method, we compute the adsorption properties for specified
thermodynamic conditions.  We take a tetragonal unit cell, of side
5.12 nm, with a hard wall at 10 nm above the surface to confine the
coexisting vapor phase without causing capillary condensation
(sufficient to contain 25 layers of benzene).  We assume periodic
boundary conditions which, although sacrificing the accuracy of the
long range QC structure, do not interfere with short-to-moderate
length scales, representative of the hydrocarbon order.  The substrate
is reproduced with an 8-layer Al-Ni-Co slab, where the atom
coordinates are derived from an experimental low-energy electron
diffraction (LEED) study \cite{Ferralis_2004b}.  The inter-molecular
interactions (adsorbate-adsorbate) are calculated as a sum of pair
interactions between atoms.  Buckingham potentials are used for
methane \cite{Tsuzuki93,Tsuzuki94} and benzene
\cite{Califano79,Chelli01} while a Morse potential is employed for
propane \cite{Jalkanen02}.  We have developed embedded-atom method
(EAM) potentials \cite{ref_EamOrigPRL} to model the many-body
intra-molecular interactions, adsorbate-substrate (C-Al, C-Co, C-Ni,
H-Al, H-Co, H-Ni), and substrate-substrate (Al-Al, Al-Co, Al-Ni,
Co-Co, Co-Ni, Ni-Ni).  The EAM embedding functions are taken as
natural cubic splines using charge density functionals from Herman
\cite{Rho}, while the EAM pair energies have Morse potential forms
\cite{Morse_pot} with Haftel's mixing scheme
\cite{Haftel_PRB_48_1993}, if necessary.  The EAM potential parameters
are fitted from {\it ab initio} energies calculated using the Vienna
Ab-initio Simulation Package (VASP) \cite{Kresse93} with
exchange-correlation functionals as parameterized by Perdew, Burke,
and Ernzerhof (PBE) \cite{PBE96} for the generalized gradient
approximation(GGA), and projector augmented-wave (PAW)
\cite{Blochl_PRB_50_1994} pseudopotentials.  The EAM potential
parameters, fitted using the {\small SIMPLEX} method
\cite{Nelder_simplex}, are summarized in the EPAPS material
\cite{EPAPS}.  Quasicrystal approximants (crystals having similar
short-range order) are used to address the periodic boundary
conditions required by the {\it ab initio} package: we use
Al$_{29}$Co$_4$Ni$_8$ (dB1), Al$_{17}$Co$_5$Ni$_3$ (dH1), and
Al$_{34}$Co$_4$Ni$_{12}$ (dH2) \cite{ref_alnicodb}.  Our calculations
show that alkanes and benzene do not dissociate on such substrates,
which do not undergo any considerable relaxation upon the adsorption
of the molecules. Thus, as a first approximation, the substrate and
the molecules can be considered rigid, although the molecules are
allowed to explore all rotational degrees of freedom \cite{Note1}.

\begin{figure}[htb]
  \vspace{-1mm}
  \centerline{\epsfig{file=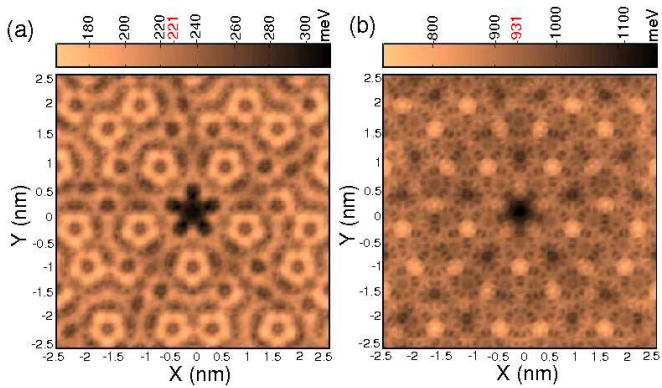,width=87mm,clip=}}
  \vspace{-3mm}
  \caption{\small (color online). Computed potential energies for
    (a) methane and (b) benzene on Al-Ni-Co, obtained by minimizing
    $V(x,y,z,\theta,\phi,\psi)$ of a single molecule with respect to
    $(z,\psi,\theta,\phi)$, variations. The average values are 221
    meV/methane and 931 meV/benzene.  }
  \vspace{-2mm}
  \label{fig1}
\end{figure}

The adsorption potential of a single molecule, $V_{min}(x,y)$,
calculated as the maximum depth as a function of normal coordinate
$(z)$ and Euler angles $(\theta,\phi,\psi)$, is both deep and
extremely corrugated.  Figure \ref{fig1} shows $V_{min}(x,y)$ for
methane and benzene: the dark spots indicate strong binding sites.
The average adsorption energy is 221 meV/methane, 374 meV/propane, and
931 meV/benzene.

The symmetry of the adsorption potentials for methane and benzene
reflect the pentagonal symmetry of the substrate, as illustrated in
Fig. \ref{fig1}. Propane follows a similar trend as methane, with
somewhat less corrugation due to its larger size.  The most attractive
adsorption positions for methane and propane \cite{EPAPS} are located
at the centers of pentagonal hollows having five Al atoms at the
vertices.  Conversely, the most attractive sites for benzene are the
Al-centered pentagons with 3 Al and 2 Ni atoms at the vertices.  These
hollow and Al-centered pentagons alternate every 36$^\circ$ around the
$z$ axis.

By simulating the adsorption of a single molecule, a general trend is
observed for the orientation of the adsorbant: the smaller the
molecule, the more variation in the rotation of the ground state.
Methane's ground state is highly degenerate, propane adsorbs with its
axis forming a small angle with respect to the substrate (the angle
varies from 0$^\circ$ to 10$^\circ$ depending on the adsorption site).
Benzene adsorbs with its plane parallel to the substrate.

\begin{figure}[htb]
  \vspace{-1mm}
  \centerline{\epsfig{file=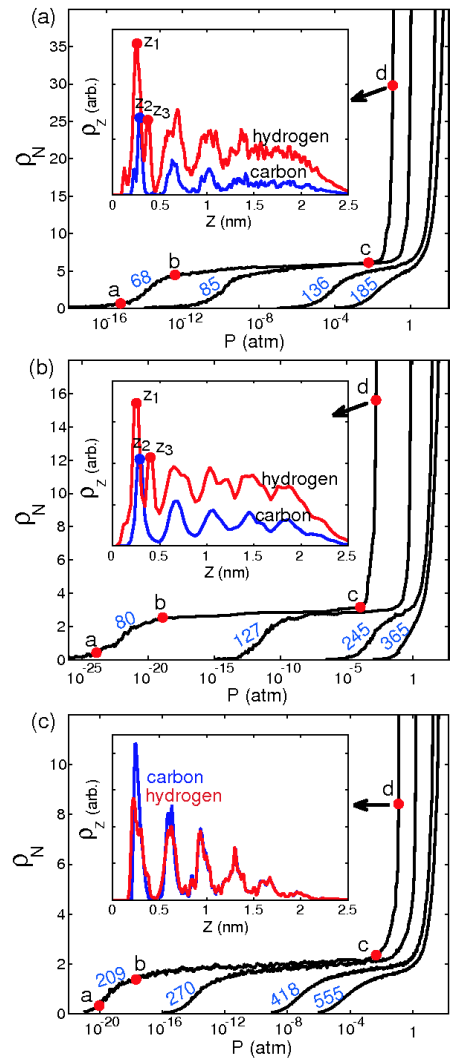,width=60mm,clip=}}
  \vspace{-3mm}
  \caption{\small (color online).  Isothermal adsorption densities
    ($\rho_N$ in molecules/nm$^2$) of hydrocarbons on a decagonal
    Al-Ni-Co: (a) methane, (b) propane, and (c) benzene. The
    simulation temperatures are reported in blue (in Kelvin).  The
    insets represent the densities along the $z$ direction at $P$
    corresponding to points ``d''.  }
  \vspace{-4mm}
  \label{fig2}
\end{figure}

Figure \ref{fig2} shows the computed adsorption isotherms $\rho_N$
(densities in molecules/nm$^2$) at different temperatures for methane
($T\hspace{-1mm}=\hspace{-1mm}68,85,136,185$ K), propane
($T\hspace{-1mm}=\hspace{-1mm}80,127,245,365$ K), and benzene
($T\hspace{-1mm}=\hspace{-1mm}209,270,418,555$ K) as functions of
pressure $P$.  The plotted quantities are the thermodynamic excess
coverages (differences between the total number of molecules and the
number that would be present if the cell were filled with uniform
vapor).  The substrate is very attractive (Fig. \ref{fig1}) hence the
hydrocarbons experience complete wetting up to the highest temperature
close to the critical temperature.  At low temperatures, the formation
of the first layers is evident from the first step in each plot (the
most left isotherms).  The formation of further layers is not clearly
observed in contrast with the observations for noble gases
\cite{XeQCPRL,XeQCPRB,XeQCJPCM,XeQCPHILMAG}.  Nevertheless, layering
in the adsorbed film is revealed by the insets of Fig. \ref{fig2}
showing the adsorption densities along the $z$ direction at the
pressures corresponding to points ``d'' of the lowest temperature
isotherms.  The relative positions of $z_1$, $z_2$, and $z_3$ indicate
that methane adsorbs mostly with three hydrogens anchored to the
substrate, while propane adsorbs mostly with five hydrogens near the
substrate (two from each end and one from the middle segment).

\begin{figure}[htb]
  \vspace{-3mm}
  \centerline{\epsfig{file=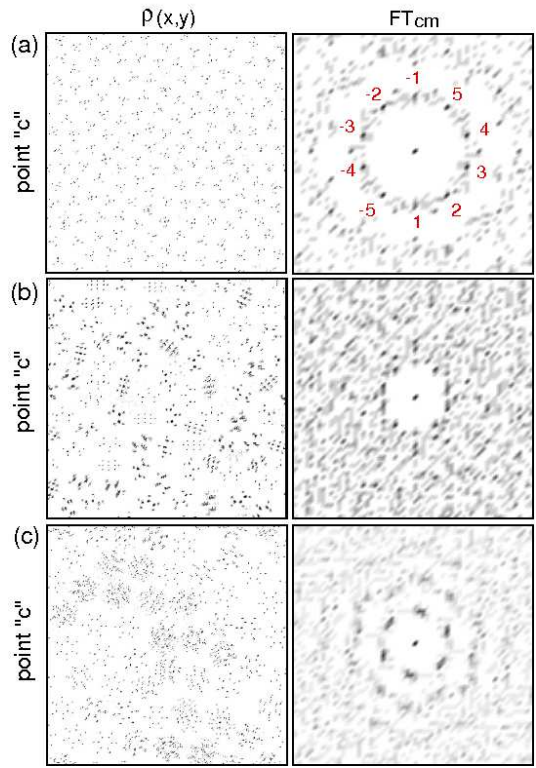,width=70mm,clip=}}
  \vspace{-3mm}
  \caption{\small Monolayer density profiles $\rho(x,y)$ and Fourier
    transforms of the density of the center of mass FT$_{cm}$ for: (a)
    methane at 68 K (point ``c'' in Figure \ref{fig2}(a)), (b) propane
    at 80 K (point ``c'' in Figure \ref{fig2}(b)), and (c) benzene at
    209 K (point ``c'' in Figure \ref{fig2}(c)) adsorbed on decagonal
    Al-Ni-Co.  }
  \vspace{-2mm}
  \label{fig3}
\end{figure}

\begin{figure}[htb]
  \centerline{\epsfig{file=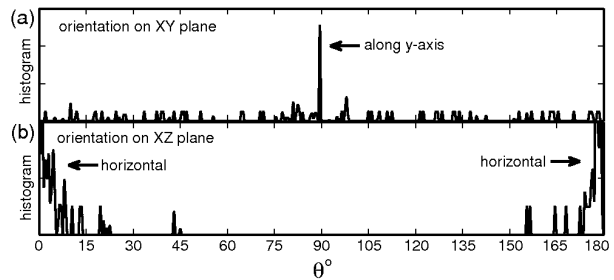,width=80mm,clip=}}
  \vspace{-3mm}
  \caption{\small Histogram of orientation of propane's axis on (a) XY
    plane and (b) XZ plane at monolayer coverage at 80 K adsorbed on
    decagonal Al-Ni-Co. The corresponding density plot is depicted in
    Figure \ref{fig3}(b).}
  \vspace{-5mm}
  \label{fig4}
\end{figure}

In the submonolayer regime (point ``a'' in Figure \ref{fig2}(a)),
methane adsorbs preferentially at the strong binding sites.  By
increasing $P$, a methane monolayer forms with pentagonal ordering
commensurate with the substrate (point ``c'' in Figure \ref{fig2}(a)).
Similar configurations are observed at all studied temperatures.  An
example of such ordering is illustrated in Fig. \ref{fig3}(a) at
$T\hspace{-1mm}=\hspace{-1mm}68$K: the density profile $\rho(x,y)$ and
the Fourier transform FT$_{cm}$ of the density of the center of mass
confirm pentagonal ordering (10 discrete spots representing 5-fold
axes in FT$_{cm}$).  Similar plots for propane at 80 K and benzene at
209 K are shown in Figure \ref{fig3} (b) and (c), respectively.  At
all simulated temperatures, benzene has 6-fold order as indicated by
its FT$_{cm}$ characteristic of triangular lattice. Unlike methane and
benzene, which adsorb in a well-defined structure, propane forms a
poorly-ordered 5-fold arrangement (clusters of molecules which form
pentagons can be seen in the EPAPS material \cite{EPAPS}). The
FT$_{cm}$ indicates a distortion involving a compression along one
direction. This is due to the interplay between the linear form of
propane molecule and the quasicrystal structure. Figure \ref{fig4}
shows a histogram of the orientations of the propane's axis projected
on the $xy$-plane (panel (a)) and the $xz$-plane (panel (b)),
corresponding to the density plot in Figure \ref{fig3}(b). Indeed,
propane adsorbs with its axis parallel with the substrate (up to
10$^\circ$) and preferentially oriented along $y$-axis.

\begin{figure}[htb]
  \vspace{-2mm}
  \centerline{\epsfig{file=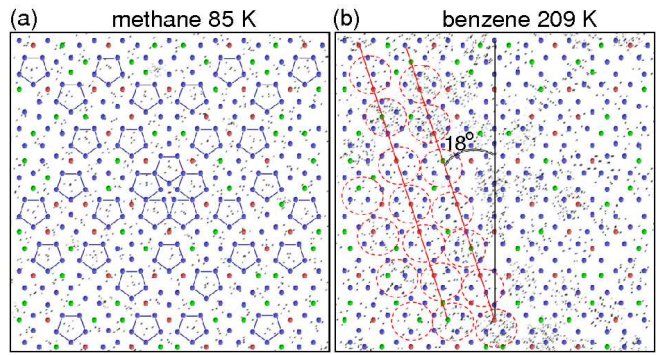,width=86mm,clip=,angle=0}}
  \vspace{-3mm}
  \caption{\small (color online). (a) methane monolayer at 85 K (b)
    benzene monolayer at 209 K adsorbed on decagonal Al-Ni-Co. Top
    layer of substrate atoms are plotted in blue (Al), green (Ni), and
    red (Co).}
  \vspace{-3mm}
  \label{fig5}
\end{figure}

Figure \ref{fig5}(a) illustrates a superposition of the methane
monolayer at 85 K and the top layer of substrate atoms. Methane
molecules are present in every hollow pentagon defined by five Al
atoms.  These pentagons, corresponding to the dominant binding sites
and depicted as dark spots in Fig. \ref{fig1}(a), are responsible of
stabilizing the quasicrystalline structure of the methane monolayer. A
similar plot for benzene at 209 K is given in Figure \ref{fig5}(b). A
domain having 6-fold order can be observed in the left part of the
figure. Dotted circles corresponding to the benzene molecules, along
with two red lines illustrate the orientation of this domain, which
makes an angle of 18$^\circ$ from the $y$-axis.  Thus, the benzene
lattice is oriented along a symmetry direction of the quasicrystal, as
observed previously for the Xe monolayer.

Previosuly, by studying adsorbed monolayer of rare gases
\cite{XeQCPRB,XeQCJPCM}, we found that the crucial parameter in
determining the overlayer structure is the relative size mismatch
between adsorbate and substrate's characteristic length.  The mismatch
is defined as $\delta_m \equiv(d_{r}-\lambda_r)/\lambda_r$ where
$d_{r}$ is the distance between rows in a 2D close-packed arrangement
of adsorbates \cite{XeQCJPCM,Note3}, and $\lambda_r=0.381$ nm is the
quasicrystal's characteristic length \cite{Ferralis_2004}.  Thus,
$\delta_m$ measures the relative mismatch between an adsorbed
$\{111\}$ closed packed plane of adsorbates and the QC surface.  The
proposed rule states that the ordering transition exists if and only
if $\delta_m>0$ \cite{XeQCJPCM}.  For methane, by starting from the
lattice parameter of the cubic cell \cite{Mooy_1931,James_1959}, we
have $\delta_m($CH$_3)\sim-0.055$.  For benzene, by using $d_{r}$
obtained from the simulation of the 6-fold phase (using a flat
substrate with the well depth of 931 meV), we get
$\delta_m($C$_6$H$_6)\sim 0.617$.  Both methane and benzene satisfy
the mismatch rule extending its applicability (the transition can not
be defined for propane since it does not form pentagonal or triangular
arrangements).

\begin{table}[tbh]
  \vspace{-5mm}
  \caption{\small Summary of adsorbed rare gases and selected
    hydrocarbons undergoing (or not) the commensurate $\rightarrow$
    incommensurate transition on-Al-Ni-Co.  iNe, iXe, and dXe are
    inflated/deflated noble gases \cite{XeQCJPCM}.  Triangles and
    pentagons indicate triangular lattice and fivefold structure,
    respectively.  }
  \vspace{-7mm}
  \begin{center}
  \item[]\begin{tabular}{cccc}
    \hline \hline
    & $\delta_m$ & transition & monolayer\\
    \hline
    methane & -0.055  &  No & $\pentagon$ \\
    benzene & 0.617 &    Yes & $\triangle$ \\
    \hline 
    Xe \cite{XeQCPRL,XeQCPRB} &  0.016 &  Yes & $\triangle$ \\
    Ne \cite{XeQCJPCM}& -0.311 &  No & $\triangle$ + $\pentagon$ \\
    Ar \cite{XeQCJPCM}& -0.158 &  No & $\triangle$ + $\pentagon$ \\
    Kr \cite{XeQCJPCM}& -0.108 &  No & $\triangle$+ $\pentagon$ \\
    iNe$^{(1)}$ \cite{XeQCJPCM} &  0.016 &  Yes & $\triangle$ \\
    dXe$^{(1)}$,dXe$^{(2)}$ \cite{XeQCJPCM}& -0.311,-0.034 &  No & $\triangle$ + $\pentagon$ \\
    iXe$^{(1)}$,iXe$^{(2)}$ \cite{XeQCJPCM}& 0.363, 0.672 &  Yes & $\triangle$ \\
     \hline \hline
  \end{tabular}
    \begin{tabular}{c}
    \end{tabular}
  \end{center}
  \label{table_mismatch} 
\end{table}
\vspace{-1mm}

Table \ref{table_mismatch} summarizes the results obtained for all
hydrocarbons and rare gases that we have studied so far.  When the
transition is not present, the monolayer does not form a structure
having long-range periodic order.  From the point of view of the
potential for superlubricity, all of the gases that are smaller than
Xe do not form periodic films.  In addition, the linear molecule
propane, which is larger than Xe, does not.  This suggests that the
shape as well as the size is an important factor in the ordering, and
it may bode well for the use of linear hydrocarbons as lubricants on
quasicyrstalline films.

Research supported by NSF (DMR-0505160, DMR-0639822) and ACS (\#PRF-45814-G5).
We thank the Teragrid Partnership (Texas Advanced Computing Center, TACC) for computational support (MCA-07S005).

\end{document}